# Data Confidentiality in Mobile Ad hoc Networks

Hamza Aldabbas, Tariq Alwada'n, Helge Janicke, Ali Al-Bayatti

Software Technology Research Laboratory (STRL), De Montfort University,
Leicester, United Kingdom
`{hamza, tariq, heljanic, alihmohd}@dmu.ac.uk`

**ABSTRACT**

*Mobile ad hoc networks (MANETs) are self-configuring infrastructure-less networks comprised of mobile nodes that communicate over wireless links without any central control on a peer-to-peer basis. These individual nodes act as routers to forward both their own data and also their neighbours' data by sending and receiving packets to and from other nodes in the network. The relatively easy configuration and the quick deployment make ad hoc networks suitable the emergency situations (such as human or natural disasters) and for military units in enemy territory. Securing data dissemination between these nodes in such networks, however, is a very challenging task. Exposing such information to anyone else other than the intended nodes could cause a privacy and confidentiality breach, particularly in military scenarios. In this paper we present a novel framework to enhance the privacy and data confidentiality in mobile ad hoc networks by attaching the originator policies to the messages as they are sent between nodes. We evaluate our framework using the Network Simulator (NS-2) to check whether the privacy and confidentiality of the originator are met. For this we implemented the Policy Enforcement Points (PEPs), as NS-2 agents that manage and enforce the policies attached to packets at every node in the MANET.*

## 1. INTRODUCTION

Mobile *ad hoc* networks are autonomous systems comprised of a number of mobile nodes that communicate using wireless transmission. They are self-organized, self-configured and self-controlled infrastructure-less networks. This kind of network has the advantage of being able to be set up and deployed quickly because it has a simple infrastructure set-up and no central administration [1]. Obvious examples are in the military or the emergency services. One scenario is establishing communication between various agents in a disaster recovery operation where e.g. fire fighters need to connect to local ambulances and traffic control in circumstances where the normal communication infrastructure is destroyed or otherwise rendered unusable. In such situations a collection of mobile nodes with wireless network interface can form a transitory network [2]. These networks are particularly useful to those mobile users who need to communicate in situations where no fixed wired infrastructures are available. However, the salient feature of creating a network 'on the fly' without requiring any prearranged infrastructure gave mobile *ad hoc* networks an appreciated interest in both industrial and military systems. The key challenges in MANETs design come from the decentralised nature, self-organisation, self-management, and also the fact that all communications are carried over wireless links in short-range communication [3]. In addition to that the topology in the network is dynamically changed because of the high mobility nature in (MANET). Therefore, all these unique characteristics present appreciable challenges for MANETs [4, 5, 6].

In comparison with wired networks where the devices must have physical access to the network medium, mobile *ad hoc* networks have no apparent secure boundary. Once the attackers are in the transmission range of any other device, then they can join and communicate with all devices in range. Indeed the nature of mobility in *ad hoc* networks, and the liberty to join, move and leave networks makes the MANET vulnerable to attacks, which can result from any device in the same transmission range [3].

Whilst devices used in wired networks get their electrical supply directly through available power grids, in MANET nodes are generally operated by small batteries with a limited lifetime. This makes nodes unable to perform intensive computations over prolonged periods of time. An attacker on the other hand is typically able to provide sufficient power-supply and thus must be assumed to be able to perform intensive computations [1], meaning that attack and defence in these networks is not levelled. The lack of centralized management in MANET makes the detection of attacks a very complicated issue. Mobile *ad hoc* networks are highly dynamic and large scale, and they cannot be easily monitored; benign (non-malignant) failures in MANETs are fairly common, e.g. transmission destructions and packet dropping. As a result, malicious failures will be more difficult to identify. Since security is an essential component in a hostile environment, these unique characteristics of mobile *ad hoc* networks raise challenges that security requirements must address [7, 8].

There has been appreciable work by the research community [7, 9, 10, 11, 12] in message encryption, digital signature, key management, etc. that address the confidentiality of messages in transit over the unsecured medium. Many challenges particularly related to the privacy and data confidentiality, however, remain to be solved. Existing approaches which have been used in MANETs such as access control, digital signature, and encryption focused only in securing the channel during the transmission, however how these nodes act after and use this information has been mostly neglected.

In this paper, we review the main security issues and existing solutions in MANET, in particular the area of security of MANET which has not been widely addressed. We present a policy-based architecture to control the data dissemination in MANETs that is build on the automatic communication of policies between nodes and draws on concepts developed for other forms of networks e.g. [28]. The purpose of this architecture is to keep data confidential not only during communication, but also after it has been transmitted to another node. Ensuring that the contents of messages are kept secret to an originator-defined subset of peers in MANETs.

In this paper, we highlight the characteristics in MANETs in Section 2, and focus on security issues in Section 3. In Section 4 we present some of the previous work on securing MANETs to which we relate our proposed policy-based architecture and the algorithm for updating policies in Section 5. We discuss advantages and limitations of our approach in section 6 and conclude the paper in Section 7 where we also summarise our findings and outline our future work in this area.

## 2. Characteristics of MANET:

Mobile *ad hoc* networks (MANET) are autonomous systems of mobile nodes connected by wireless links. These nodes are therefore free to move arbitrarily; thus, the topology of wireless networks can be changed dynamically and unpredictably. MANETs have many characteristics that make them are different from other wireless and wired networks that are well recognised [13, 14, 15, 16, 8]:

1. **Multi-hop communications**: The communication in MANET between any two remote nodes is performed by numerous intermediary nodes whose functions are to relay data-packets from one point to another. Thus, *ad hoc* network requires the support of multi-hop communications. For example, in Figure 1, nodes A and D must engage the help of nodes B and C to relay data-packets between them in order to communicate.

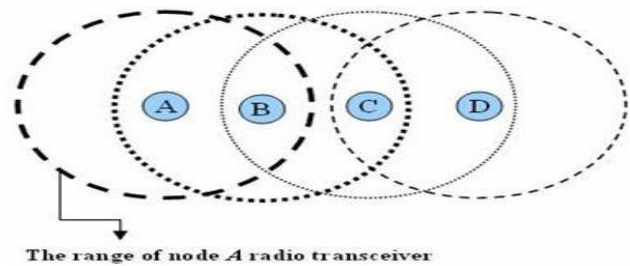

Figure 1: MANET of four nodes and their transmission ranges

2. **Constrained Resources:** Generally, most MANET devices are small hand-held devices ranging from personal digital assistants (PDAs) and laptops down to cell phones. These devices indeed have limitations because of their restricted nature; they are often battery-operated, with small processing and storage facilities.

3. **Infrastructure less:** MANETs are formed based on the collaboration between autonomous nodes, peer-to-peer nodes that need to communicate with each other for special purpose, without any pre-planned or base station.

4. **Dynamic Topology:** MANET nodes are free to move, hence the connectivity between nodes in MANET can change with time, because nodes can move arbitrarily; thus the nodes can be dynamically inside and outside the network, constantly changing their links and topology, leading to change in the routing information all the time due to the movement of the nodes. Therefore, the communicated links between nodes in MANET can be bi-directional or unidirectional.

5. **Limited Device Security:** MANETs devices are usually small and can be transported from one place to another, then they are not constrained by location. Unfortunately, as a result these devices could be easily lost, stolen or damaged.

6. **Limited Physical Security:** Generally, MANETs are more susceptible to physical layer's attacks than wired network; the possibility of spoofing, eavesdropping, jamming and denial of service (DoS) attacks should be carefully considered. By contrast the decentralised nature of MANET makes them better protected against single failure points.

7. **Short Range Connectivity:** MANETs rely on radio frequency (RF) technology to connect, which is in general considered to be short range communication. For that reason, the nodes that want to communicate directly need to be in the close frequency range of each other. In order to deal with this limitation, multi-hop routing mechanisms have therefore to be used to connect distant nodes through intermediary ones that operate as routers.

## 3. Network Security Considerations

Since security is an essential component in MANET, the striking features of mobile *ad hoc* networks raise both challenges and opportunities in achieving these security goals, Unlike other traditional networks (wired) where nodes must have physical access to the network or communicate through several defence perimeters like firewalls and gateways, MANET uses the wireless medium so attacks on a wireless network can come from all directions and target any node. This provides a larger surface of attack ranging from passive attacks, such as "tapping" to active attacks, such as message replay, message leakage, contamination and distortion. This means that a MANET does not have a clear line of defence, and every node must be prepared to defend against the different kind of attacks [17].

MANETs present a new set of challenges to security design as security solutions must achieve both data protection and efficient network performance [18]. While we addressing network security, we have to consider the security requirements to take account of the functionality required to provide a secure networking system.

## 3.1. Security Requirements

The security requirements specified below specified by International Telecommunications Union (ITU-T) represented in their recommendation X.805 and X.800 [19, 4, 18, 29]:

1. **Authentication:** Authentication is essential to verify the identity of every node in MANET and its eligibility to access the network. This means that, nodes in MANETs are required to verify the identities of the communicated entities in the network, to make sure that these nodes are communicating with the correct entity.

2. **Authorisation and Access Control:** Each node in MANET is required to have the access to shared resources, services and personal information on the network. In addition, nodes should be capable of restricting each other from accessing their private information. There are many techniques that can be used for access control such as Discretionary Access Control (DAC), Mandatory Access Control (MAC) and Role Based Access Control (RBAC).

3. **Privacy and confidentiality:** Each node has to secure both the information that is exchanged between each other; and secure the location information and the data stored on these nodes. Privacy means preventing the identity and the location of the nodes from being disclosed to any other entities, while confidentiality means keeping the secrecy of the exchanged data from being revealed to those who have not permission to access it.

4. **Availability and survivability:** The network services and applications in MANET should be accessible, when needed, even in the presence of faults or malicious attack such as denial-of-service attack (DoS). While survivability means the capability of the network to restore its normal services under such these conditions. These two requirements should be supported in MANET.

5. **Data integrity:** The data transmitted between nodes in MANET should be received to the intended entities without been tampered with or changed by unauthorised modification. This requirement is essential especially in military, banking and aircraft control systems, where data modification would make potential damage.

6. **Non-repudiation:** This ensures that nodes in MANET when sending or receiving data-packets should not be able to deny their responsibilities of those actions. This requirement is essential especially when disputes are investigated to determine the misbehaved entity. Therefore digital signature technique is used to achieve this requirement to prove that the message was received from or sent by the alleged node.

## 4. State Of The Art

Existing approaches in security which have been applied to MANETs are for example using traditional cryptographic solutions based on public key certificates [24] to maintain trust, in which a Trusted Third Party (TTP) or Certificate Authority (CA) [25] certifies the identity associated with a public key of each communicated entities. Iman Almomani and Hussein Zedan [26] proposed a comprehensive, top-down, end-to-end security solution for MANET based upon a well defined architecture and exploiting two of the ITU-T recommendations: X.800, and X.805. Such approaches can therefore provide end-to-end secure communication channels and are mainly focused on message confidentiality, integrity and non-repudiation, they do not consider however the trust management of the communicated entities, and how these certified entities act is left to the application layer [20]. Ali Al-Bayatti *et al* [27] proposed behaviour detection algorithm combined with threshold cryptography digital certificates to satisfy prevention and detection to securely manage Mobile *Ad hoc* Network of Networks (MANoNs). Lidong Zhou *et al* [10] studied the security threats, vulnerabilities and challenges which faces the *ad hoc* network. In their work [10] they protected the packets sent between nodes by choosing the secure routing path to the destination node based on the redundancies routes between nodes to maintain the availability requirement. This is because all key-based cryptographic approaches such as digital signature need a proper and secure key management scheme to bind between the public and private keys to the nodes in the network; Lidong Zhou used replication and a new cryptographic technique (threshold cryptography) [21, 22] to build a secure key management process to achieve trust between a set of servers in *ad hoc* networks by distributing trust among aggregation of nodes to certify that these nodes are trustworthy.

Securing the routing in mobile *ad hoc* network (MANET) has also been given much attention by the researchers; many approaches, therefore, have been proposed to deal with external attack. Sirios and Kent [23] proposed an approach to protect the packet sent to multi receivers by using keyed one-way hash function supported by windowed sequence number to ensure data integrity. The trust issue in communication systems as in mobile *ad hoc* network is a challenging task to achieve. The framework that we are proposing has some relations with secure routing protocols, as it determines the sharing of information, however we consider policies to operate at a higher level in the protocol stack where application specific trust decisions can be made. Public Key Infrastructure (PKI) and cryptography are achieving a kind of a quasi-trust before the communication is started. However, how the nodes act after that is a controversial issue as untrusted nodes cannot be predicted without establishing tracing techniques to ensure that they are not misbehaving whilst participating in the MANET.

In commercial and medical contexts, individuals should be able to keep and manage access to their personal information by choosing to which entities information should be disclosed in a discretionary way. Individuals expect that their personal data, personal information, or Personally identifiable information (PII) to be securely communicated and shred between different organisations such as their names, addresses, phone numbers, national insurance numbers, credit card details, passwords, or date of birth (DOB). Therefore, Siani Pearson and Marco Casassa Mont [28] employed a clever idea of sticking policies with data to control how the PII should be processed, handled, shared with other parties.

## 5. Our Proposed Framework

In this paper, we show how revealing secret information by a malicious node (inside the network) to unwanted nodes could cause a privacy and confidentiality breach. Therefore traditional encryption tools being commonly used solve one part of the problem by encrypting data exchanged between nodes using the public key of the destination node and then decrypting the packet by the destination's private key but how the destination behave after is neglected. Using a mechanism which employs access control to ensure confidentiality is a real possibility, so our work intends to use this mechanism, in particular Discretionary Access Control (DAC) to ensure data confidentiality and privacy of the originator node in MANETs.

### 5.1. Scenario

To motivate our approach we present a scenario in which a close military alliance between two countries wants to share tactical mission information only between themselves and not with other coalition members.
Considering three nodes A,B,C in Figure 2, where node A and B are allocated to British and US armies, and C is allocated to Afghan army. Node A wants to send a tactical message for the mission that says "'we are going to start the mission after 8 pm"' to node B, however node A does not want node B send the message to node C because node C is not trusted by A. How can node A trust node B not to send the message to node C?

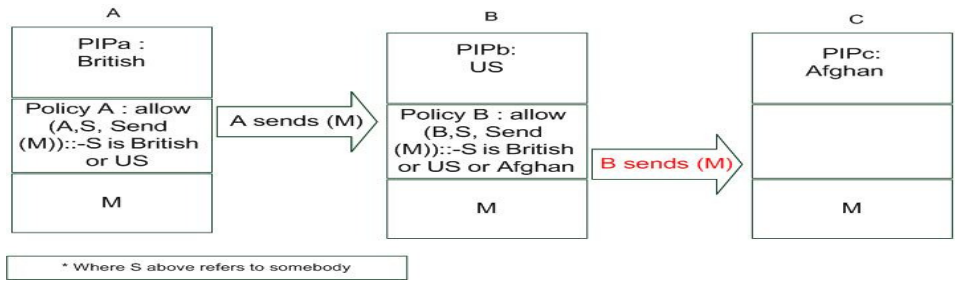

Figure 2: node B disclose the message to C

Node A sends the message (M) to node B, node B now knows the message (M). However depend on its policy node B can send the message (M) and disclose it to node C. Which it is the problem of the node A privacy.

The goal of our proposed approach is to solve this problem by allowing the originator to specify a high level policy which will automatically apply and enforce itself to all the communicated entities on the network. This is done by attaching the policy of the originator (A) with the message (M) to control the access to it, which is capable to define who are allowed to access that message. In this way the policy of node A attached with the message (M), tells node B to which node can the message (M) be send to (only British or US armies can receive the message) as in Figure 3 :

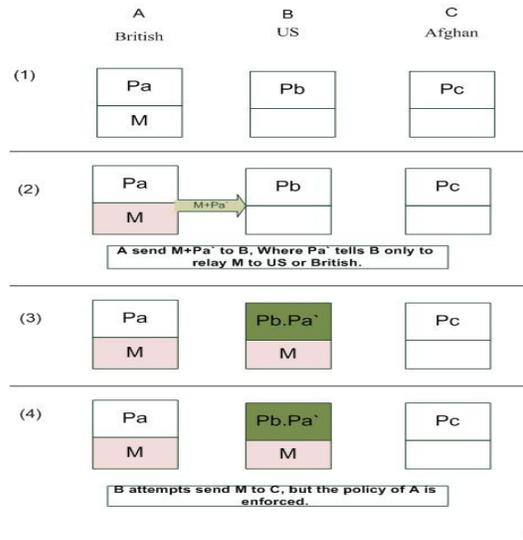

Figure 3: Prevention Node B Of Disclosing M to node C

node A sends the message (M) + policy of node A which tells node B to allow sending the message (M) to any node if its British or US nodes. Here after node A sent the message (M) to node B attached by the node A policy. node B receives the packet and now knows the message (M) in addition of that it knows the policy of A:
*Allow (Node B, Send(M) to S: if S is British or US, where S refers to somebody).*

## 5.2 Our Framework

Figure 4 presents the proposed framework, where policies are used to enforce access control to such information sent by the originator to other entities in the system, our framework will be introduced in every entity in the communicated systems.

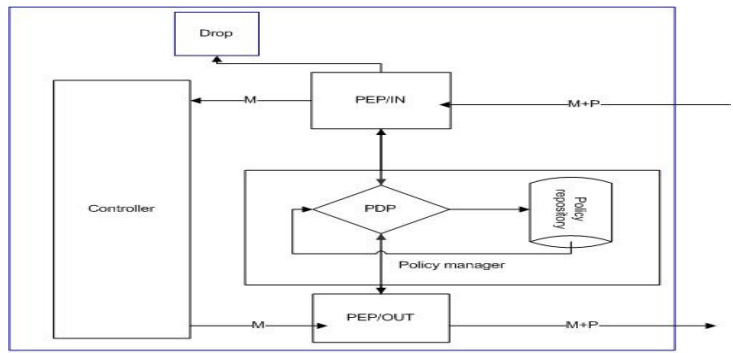

Figure 4: The proposed framework

Our framework is composed of four components as they shown in the Figure 4:
1. Policy Enforcement Point (PEP/OUT): executes and enforces policy decisions in the sender node, this component installed at the transmitter interface that does merge system's policy with the message sent to others systems. In our simulation we configured the send function to achieve the functionalities of this component.
2. Policy Enforcement Point (PEP/IN): executes and enforces policy decisions in the receiver node, this component installed at the receiving interface that does inverse process at the receptive system, splitting and dividing the message from the policy attached. In our simulation we configured the receive function to achieve the functionalities of this component.
3. Policy Decision Point (PDP): plays a crucial role in both the sender and the receiver side in our framework, and helping other components to do their jobs. In our simulation we configured this function as a shared source code to achieve the functionalities of this component in all nodes.
4. Controller that process and store the information received from the other components.

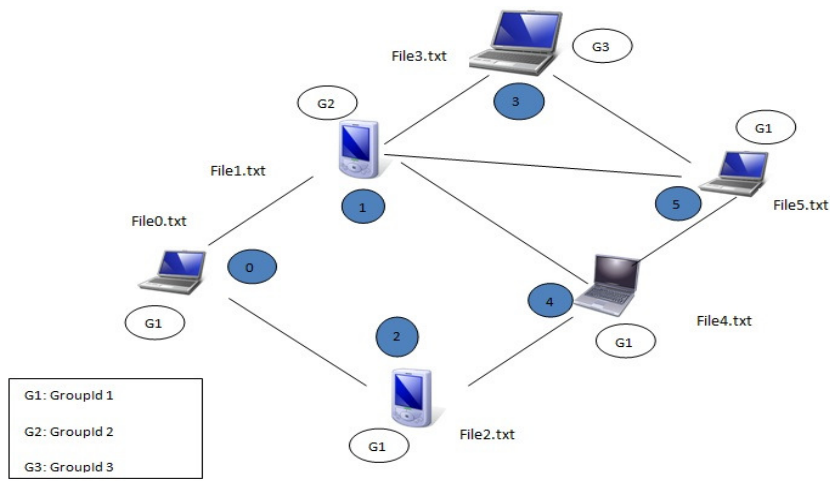

Figure 5: Example illustrating the Algorithm chart

In Figure 5 we show an example of six nodes, assuming that each node in the system has a group-id number, means we are classifying the nodes in our work into different groups, which in our case three groups: G1, G2, and G3. The first group has node 0, node 2, node 4 and node 5. Whereas node 1, node 3 are in G2 and G3 respectively. In our work we make node 0 broadcast a message to all nodes in the group-id specified in the policy file at file0.txt in node node 0, and we call this group-id in this situation a permitted group as shown in the algorithm chart in Figure 6. If file0.txt as in the example has been marked with the group-id 1 that means

231

only nodes in the group G1 can receive the packet. Node 0 will start searching for the adjacent nodes in transmission range. In this example node 0 will find node 1 and node 2 and select them as potential destinations (dest1=n1, dest2=n2). Now node 0 will check if dest1 in the permitted group or not, and do the same to dest2 also. In the algorithm (Figure 6) this depicted as Getgroupid (dest) process and checks if the group-id is permitted. In this example it will be permitted for node 2 as it is in the G1. Node 0 will send to node 2 not only the packet it also sends its policy contained in file0.txt, whereas node 2 will create a packet handler to receive the packet. Once it received the policy, node 2 will be updated according to received policy updating its old policy.

Now, when node 2 at another time wants to broadcast the message again will start and do the same process like node 0. This time, however, node 2 will send to node 4 but not to node 0 because node 0 is the originator of the packet as shown in the algorithm chart in Figure 6, and the system will continue through the same steps for other nodes.

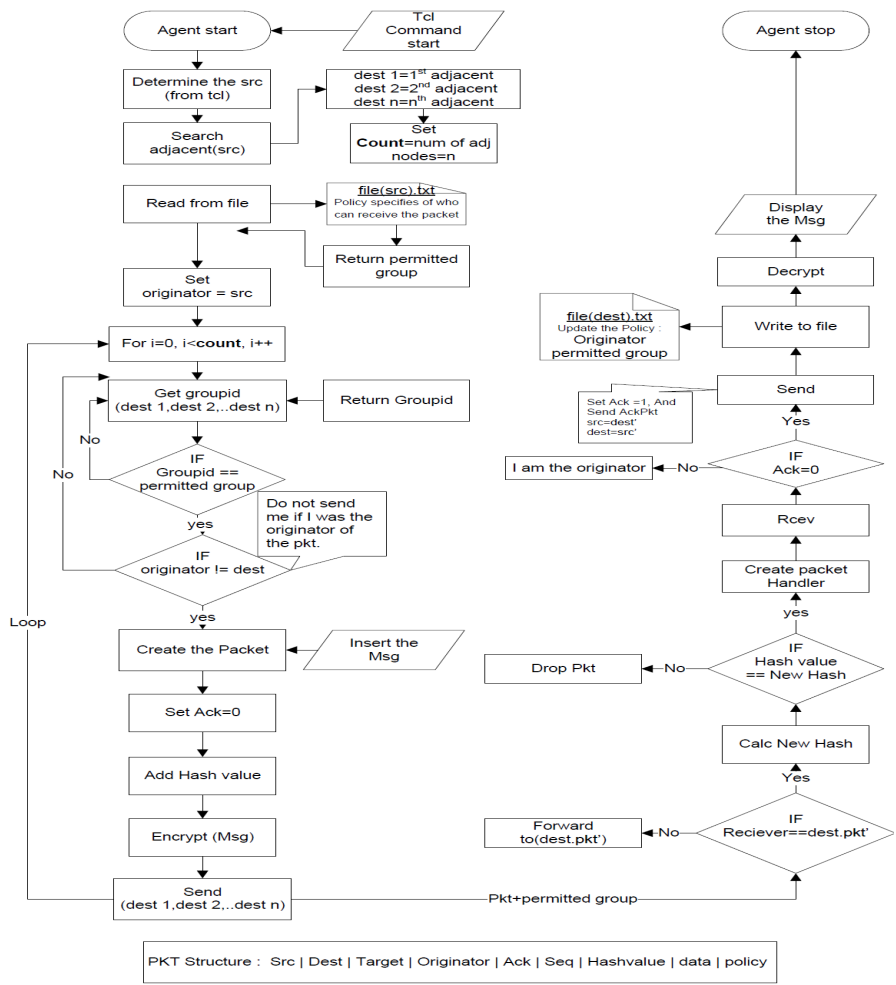

Figure 6: Our Chart Algorithm and packet structure

## 6. Result and Discussion

In this work we used the Network Simulator (NS-2) which is a real network environment simulator, which showed only intended nodes can receive the packet which has been sent by the source. We simulated our approach into multi variable number of nodes where the originator node disseminates the packet (Data+Policy) to the other nodes. Our result from the tracing file and the nam showed that only nodes in the permitted group-id can receive the packet because of the restriction which has issued from the originator node 'not to send the packet to nodes in different group-id.

We assumed that all nodes in the system are trusted to enforce the policy attached within the packet, and the encryption, digital signature, and the keys management have already been done securely. So in this work we address the stage after the processes mentioned above. So implicitly that means that if a node is trusted to receive a certain information (in the clear) our framework assumes that it also will be trusted to protect this information. In the domains discussed above this appears to be a reasonable assumption that however does not protect against malicious nodes in the network that have infiltrated the system and (wrongly) gained the trust of message originators. Similarly the system does not prevent out-of-band communications and assumes the data is communicated using the provided infrastructure. The protection here is that automated services that provide e.g. situational awareness are trust-enabled to limit the dissemination of information.

We simulated our policy-based agent with a variable number of UDP agents simultaneously to check what happen if all agents are started in the simulation and how the time necessary a packet to be transmitted across a network from source to destination will be affected. In Figure 7 we measured the delay time versus number of cbr traffics which are depicted on the y-axis and x-axis respectively. The result of this figure showed that as the number of cbr traffic increases, the delay time of both agents will increase. We started with 1 cbr traffic, 2, 3 and 4 with and without our policy-based agent to be started at different sources and destinations to measure the average of the delay time between them.

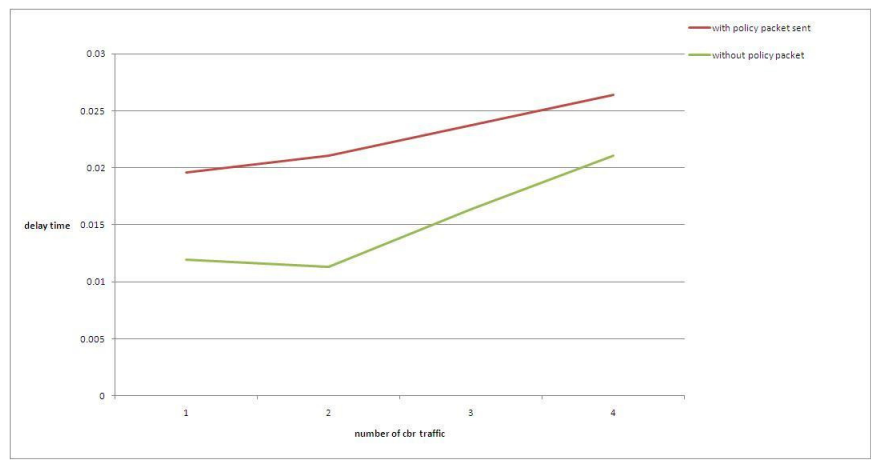

Figure 7: Delay time

It is clear from Figure 7. that there is a delay introduced by sending additional information that contains the policies, but not to a degree that would yield the system unusable.

## 7. Conclusion

In this paper we conclude that our framework achieves the source policy to send the packet for intended nodes only in the network, on top of that we highlighted the special considerations for security in MANETs and provided an extensive overview of related work and the state of the art in this area. To our knowledge, none of the related work addressed the issue of controlling the information flow in MANETs. We presented a scenario drawn from the military domain, where the impact of this form of confidentiality breach is evident and a real risk. We provided an architecture that addresses this problem by automatically attaching policies to the messages that identify how the information can be used by the receiver, thus limiting the relay of messages based on the originator's confidentiality requirements. We currently assume that all nodes in our system are trusted to correctly enforce the policies that are attached to the message and provide a communication system that includes a policy processing layer dealing with the merging of existing and received policies. However, the assumption that all are trusted is strong. In future work we shall relax this assumption by providing traceability, *viz.* water-marking messages in such a way that they are identified as compromised, they will then lead to a dynamic adjustment of the originator policy based on the resulting trust.

## Authors:

### Hamza Aldabbas

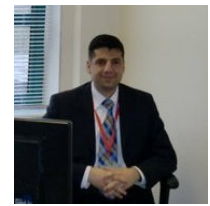

I am currently a PhD student at STRL (Software Technology Research Laboratory), De Montfort University, Leicester - United Kingdom. Obtained M.Sc degree in computer science from Al-Balqa'a Applied University-Jordan (2006—2009). B.Sc. degree in Computer Information System from Al-Balqa'a Applied University- Jordan   (2002—2006).  I am also a part time lecturer at De Montfort University, involved in teaching and project supervision at B.Sc. & M.Sc levels (2010-until now).

## Dr. Helge Janicke

I am working as a Senior Lecturer in Computer Security here at De Montfort University. I am located in the Software Technology Research Laboratory (STRL) which is part of the Faculty of Technology. My main research interests are in Computer Security, Policy Based Management, Formal Methods and Software Quality Assurance.

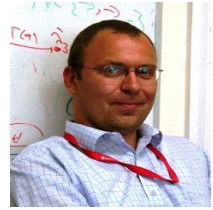

## Dr. Ali Hilal Al-Bayatti

I am currently a Research Fellow at STRL, De Montfort University, Leicester - United Kingdom. My Main research interests are in Mobile Ad hoc Networks, Vehicular Ad hoc Networks and Context-aware Systems. My main duties are to supervise Ph.D. & M.Sc students. Already, acted as an examiner board member (Internal Examiner) to four PhD students, (External Examiner) in university of Bradfordshire.

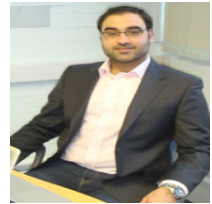

## Tariq Alwada'n

I am currently a PhD student at STRL (Software Technology Research Laboratory), De Montfort University, Leicester - United Kingdom. Obtained M.Sc degree in computer and information networks from Essex University, B.Sc in computer engineering from Al-Balqa'a Applied University-Jordan .